\documentclass[8.5pt,twoside,twocolumn]{article}
\usepackage[super,sort&compress,comma]{natbib} 
\usepackage{mhchem}
 \usepackage{times}
\usepackage{color}
\usepackage{sectsty}
\usepackage{balance} 

\usepackage{graphicx} 
\usepackage{lastpage}
\usepackage[format=plain,justification=raggedright,singlelinecheck=false,font=small,labelfont=bf,labelsep=space]{caption} 
\usepackage{fancyhdr}
\begin{document}

\thispagestyle{plain}
\fancypagestyle{plain}{
\fancyhead[L]{\includegraphics[height=8pt]{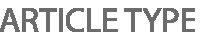}}
\fancyhead[C]{\hspace{-1cm}\includegraphics[height=20pt]{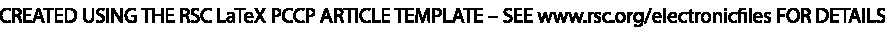}}
\fancyhead[R]{\includegraphics[height=10pt]{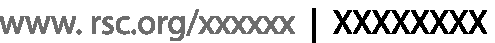}\vspace{-0.2cm}}
\renewcommand{\headrulewidth}{1pt}}
\renewcommand{\thefootnote}{\fnsymbol{footnote}}
\renewcommand\footnoterule{\vspace*{1pt}%
\hrule width 3.4in height 0.4pt \vspace*{5pt}} 
\setcounter{secnumdepth}{5}

\makeatletter 
\def\subsubsection{\@startsection{subsubsection}{3}{10pt}{-1.25ex plus -1ex minus -.1ex}{0ex plus 0ex}{\normalsize\bf}} 
\def\paragraph{\@startsection{paragraph}{4}{10pt}{-1.25ex plus -1ex minus -.1ex}{0ex plus 0ex}{\normalsize\textit}} 
\renewcommand\@biblabel[1]{#1}            
\renewcommand\@makefntext[1]%
{\noindent\makebox[0pt][r]{\@thefnmark\,}#1}
\makeatother 
\renewcommand{\figurename}{\small{Fig.}~}
\sectionfont{\large}
\subsectionfont{\normalsize} 

\fancyfoot{}
\fancyfoot[LO,RE]{\vspace{-7pt}\includegraphics[height=9pt]{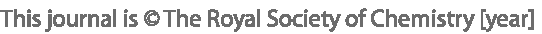}}
\fancyfoot[CO]{\vspace{-7.2pt}\hspace{12.2cm}\includegraphics{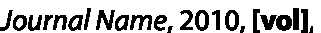}}
\fancyfoot[CE]{\vspace{-7.5pt}\hspace{-13.5cm}\includegraphics{RF}}
\fancyfoot[RO]{\footnotesize{\sffamily{1--\pageref{LastPage} ~\textbar  \hspace{2pt}\thepage}}}
\fancyfoot[LE]{\footnotesize{\sffamily{\thepage~\textbar\hspace{3.45cm} 1--\pageref{LastPage}}}}
\fancyhead{}
\renewcommand{\headrulewidth}{1pt} 
\renewcommand{\footrulewidth}{1pt}
\setlength{\arrayrulewidth}{1pt}
\setlength{\columnsep}{6.5mm}
\setlength\bibsep{1pt}

\twocolumn[
  \begin{@twocolumnfalse}
\noindent\LARGE{\textbf{Anisotropic effect on dynamics of block copolymers in lamellar phases: Relaxation and the grain boundary motion}}
\vspace{0.6cm}

\noindent\large{\textbf{Chi-Deuk Yoo,$^{\ast}$ and Jorge Vi\~nals}}\vspace{0.5cm}

\noindent\textit{\small{\textbf{Received Xth XXXXXXXXXX 20XX, Accepted Xth XXXXXXXXX 20XX\newline
First published on the web Xth XXXXXXXXXX 200X}}}

\noindent \textbf{\small{DOI: 10.1039/b000000x}}
\vspace{0.6cm}

\noindent \normalsize{
We consider the effects of anisotropic diffusion and hydrodynamic flows on the relaxation time scales of the lamellar phase of a diblock copolymer. We first extend the two-fluid model of a polymer solution to a block copolymer, and include a tensor mobility for the diffusive relaxation of monomer composition which is consistent with the uniaxial symmetry of the lamellar phase. The resulting equation is coupled to the momentum conservation equation, allowing also for a dissipative stress tensor for a uniaxial fluid. We then study the linear relaxation of weakly perturbed lamellae, and the motion of a tilt grain boundary separating two semi-infinite domains. We find that anisotropic diffusion has a negligible effect on the linear relaxation of the layered phase (in the long wavelenght limit), whereas the introduction of hydrodynamic flows considerably speeds the decay to a rate proportional to $Q^{2}$, where $Q\ll 1$ is the wavenumber of a transverse perturbation to the lamellar phase (diffusive relaxation scales as $Q^{4}$ instead). On the other hand, grain boundary motion is siginificantly affected by anisotropic diffusion because of the coupling between undulation and permeation diffusive modes within the grain boundary region.}
\vspace{0.5cm}
 \end{@twocolumnfalse}
  ]

\section{Introduction}


\footnotetext{\textit{School of Physics and Astronomy, and Minnesota
Supercomputing Institute, University of Minnesota, 116 Church Street S.E., Minneapolis, MN 55455, USA. E-mail: yoo@physics.umn.edu}}



Block copolymers are finding numerous applications in nanotechnology \cite{re:park03}, and have been of great interest in soft matter science because they undergo microphase separation to ordered phases of different symmetries (For a brief review see Ref.~\cite{bates99} and references therein). Order parameter models to describe equilibrium properties and the microphase phase diagram were given in Refs.~\cite{leibler80, ohta86, fredrickson87}. Microphase separation also brings about interesting dynamical properties, including unusual rheological response \cite{rosedale90, larson93, patel95, re:wu05} and orientation selection during shear aligning. \cite{koppi92, koppi93}

In principle, the low-frequency and long-wavelength hydrodynamic equations of motion for ordered systems can be derived by considering conservation laws and symmetry arguments. \cite{martin72, forster.hydro.75} For a diblock copolymer there are six conservation laws: Total mass of each monomer, three components of momentum, and energy. 
The two-fluid model has been widely used to describe the dynamical behavior of polymer solutions and blends in this hydrodynamic regime \cite{onuki90, doi92, milner93}. More recently, the two-fluid model has been further extended to describe the dynamics of block copolymer melts \cite{hall06, ceniceros09}. The resulting governing equations involve dissipative dynamics for the order parameter that represents local monomer composition changes, and overdamped (or Stokesian) dynamics for the velocity fields. However, both dissipative relaxation of the order parameter and hydrodynamic equations assume that the block copolymer phase is an isotropic fluid. For example, the order parameter equation contains a current associated with free energy dissipation due to the relative motion of the two types of monomers, with a kinetic coefficient (or mobility) that does not respect the uniaxial symmetry of the lamellar phase.

We address here the constitutive relations for a block copolymer lamellar phase that reflect the uniaxial symmetry of this ordered phase, and derive the corresponding equations of motion in the hydrodynamic regime (Sec. II). This requires both an anisotropic mobility tensor for order parameter diffusion, and anisotropic viscosities for the dissipative stress tensor. We next investigate in Sec. III the linear relaxation of weakly perturbed lamellae, and in Sec. IV the motion of a tilt grain boundary. 

\section{Model equations}

The equations governing the dynamics of an AB-diblock copolymer in its lamellar phase can be derived by using conservation laws and broken symmetry arguments \cite{martin72, forster.hydro.75, zhang10}. According to the two-fluid model for polymer solutions, blends, or diblock copolymers \cite{onuki90, doi92, milner93} one introduces two continuity equations for each monomer
\begin{equation}
\partial_t \rho_\text{A} = - \partial_i (\rho_\text{A} v_{i \text{A}}),
\end{equation}
\begin{equation}
\partial_t \rho_\text{B} = - \partial_i (\rho_\text{B} v_{i \text{B}}),
\end{equation}
where $\rho_\text{A, B}$ and $v_{i \text{A, B}}$ are the monomer number fraction and the corresponding
%
%
velocities with A and B denoting polymers and solvent for polymer solutions or two types of monomers in polymer blends or diblock copolymers. It is customary to introduce an order parameter $\psi=\rho_\text{A} - \rho_\text{B}$ so that
\begin{equation}
\partial_t \psi + \partial_i J_i = 0,
\label{order.parameter.cont.eqn}
\end{equation}
where the flux $J_i$ has both a reversible part $J^\text{R}_i$ that accounts for advection of $\psi$, and a dissipative part $J^\text{D}_i$ responsible for energy dissipated due to the relative motion of the two types of monomers. 
By taking the difference of the continuity equations, the order parameter equation Eq.~(\ref{order.parameter.cont.eqn}) contains a reversible current $J_i^\text{R} = \psi v_i$ where the average flow velocity is $v_i = \rho_\text{A} v_{i \text{A}} + \rho_\text{B} v_{i \text{B}}$ with $\rho_\text{A} + \rho_\text{B} = 1$, and the dissipative current
%
\begin{equation}
J_i^\text{D} = 2 \rho_\text{A} \rho_\text{B} \bigg[ v_{i \text{A}} - v_{i \text{B}} \bigg].
\end{equation}
According to the two-fluid model, the relative velocity between two monomers in $J_i^\text{D}$ is obtained by employing a Rayleigh's variational principle \cite{landau.stat.phys.80} in which a dissipative function is introduced that contains the square of the relative velocity times an isotropic friction coefficient $\zeta$. Under this assumption, the resulting dissipative current is \cite{onuki90, doi92, milner93, hall06, ceniceros09}
\begin{equation}
J_i^\text{D} = - \zeta^{-1} ( \rho_\text{A} \rho_\text{B} )^2 \partial_i \mu,
\end{equation}
where $\mu= \mu_\text{A} - \mu_\text{B}$ is the difference in the monomer chemical potentials.
The constitutive law, however, must reflect the ordered phase's symmetry. In general, one would write
\begin{equation}
J_i^\text{D} = - \Lambda_{ij} \partial_j \mu,
\label{dissipative.current}
\end{equation}
with $\Lambda_{ij}$ an anisotropic kinetic constant tensor. The number of independent components of $\Lambda_{ij}$ is determined by the symmetry of the phase. For layered systems of uniaxial symmetry, there are only two independent components,
\begin{equation}
\Lambda_{ij} = \Lambda_L n_i n_j + \Lambda_T (\delta_{ij} - n_i n_j),
\end{equation}
where $n_i$ is the unit normal to the layers. In the hydrodynamic limit, and when lamellae are weakly perturbed, the unit vector $n_i$ varies slowly. We assume that in this limit $\Lambda_{ij}$ retains the same form as above with the local principal axis defined by the local normal to the disturbed layers. Furthermore, one would expect, and this is confirmed experimentally, that since chain mobility across the layers is suppressed relative to motion parallel to the layers because of large entropic barriers for copolymers to move across layers, \cite{lodge95} then $\Lambda_L \le \Lambda_T$.

For lamellar diblock copolymers in the weak segregation limit the relative chemical potential is obtained by taking the functional derivative with respect to $\psi$ of the free energy functional given by \cite{leibler80, ohta86, fredrickson87}
\begin{equation}
F = \frac{1}{2} \int d^3 r \Bigg\{
- K \psi^2 + \frac{\chi}{2} \psi^4 + \xi\bigg[ (\nabla^2 + q_0^2) \psi \bigg]^2
\Bigg\},
\end{equation}
where $q_0$ is the wavenumber of the layers, and $K$, $\chi$, and $\xi$ are coefficients that depend on the material properties of the block copolymer.
Since the reversible current $J_i^\text{R}=\psi v_i$, and with Eq.~(\ref{dissipative.current}) for $J_i^\text{D}$, the order parameter equation Eq.~(\ref{order.parameter.cont.eqn}) becomes
\begin{equation}
\partial_t \psi + \partial_i \left( \psi v_i \right) = \partial_i \bigg[ \Lambda_{ij}  \partial_j \left( \frac{\delta F}{\delta \psi} \right) \bigg].
\label{order.parameter.eqn.unscale}
\end{equation}

Given the high viscosity of block copolymer melts, the momentum conservation equation is considered in the overdamped limit (small Reynold number) \cite{zhang10}
\begin{equation}
\partial_i P - \partial_j \sigma_{ij}^\text{R}  - \partial_j \sigma_{ij}^\text{D} = 0,
\label{momentum.conservation.eqn1}
\end{equation}
which has an implicit dependence on the velocity. In this equation
$P$ is the pressure, $\sigma_{ij}^\text{R}$ the reversible elastic stress tensor, and $\sigma_{ij}^\text{D}$ the dissipative stress tensor. The gradient of the reversible elastic stress tensor is simply \cite{re:gurtin96,re:jasnow96,zhang10}
\begin{equation}
\partial_j \sigma_{ij}^\text{R} = - \psi \partial_i \left( \frac{\delta F}{\delta \psi} \right).
\end{equation}

For systems with uniaxial symmetry there are five independent viscosities in $\sigma_{ij}^\text{D}$ \cite{ericksen59}, although the number reduces to three under the assumption of incompressibility. Hence the dissipative stress tensor for an incompressible system with uniaxial symmetry can be written as \cite{ericksen59}
\begin{equation}
\sigma_{ij}^\text{D} = \alpha_1 n_i n_j n_k n_l v_{kl} + \alpha_4 v_{ij} +\alpha_{56} n_k ( n_i v_{kj} + n_j v_{ki}),
\label{viscous.stress.tensor.unscale}
\end{equation}
where the strain rate tensor is $v_{ij} = (\partial_i v_j + \partial_j v_i)/2$. Again, we will assume that for slowly varying lamellar phases, the expression (\ref{viscous.stress.tensor.unscale}) holds locally.

Finally, block copolymers are normally assumed to be incompressible fluids $\partial_i v_i=0$. Consequently, we have Eqs.~(\ref{order.parameter.eqn.unscale})-(\ref{viscous.stress.tensor.unscale}) plus the incompressibility condition $\partial_i v_i = 0$ as the governing equations for the evolution the lamellar phase of a diblock copolymer.

Before we proceed any further,we recast the governing equations of motion, Eqs.~(\ref{order.parameter.eqn.unscale})-(\ref{viscous.stress.tensor.unscale}), in terms of dimensionless quantities ${\bf x}^\prime = q_0 {\bf x}$, $t^\prime = \xi\Lambda_L q_0^6 t$, $\psi^\prime = \psi / \sqrt{\xi/\chi}q_0^2$, and $F^\prime = F/(\xi^2 q_0^5/\chi)$. With the newly defined variables the order parameter equation can be rewritten as
\begin{equation}
\partial_t \psi + v_i \partial_i \psi = \partial_i \bigg[ \Lambda_{ij}  \partial_j \left( \frac{\delta F}{\delta \psi} \right) \bigg],
\label{order.parameter.eqn}
\end{equation}
with the free energy being
\begin{equation}
F = \frac{1}{2} \int d^3 r \Bigg\{
- \epsilon \psi^2 + \frac{1}{2} \psi^4 + \bigg[ (\nabla^2 + q_0^2) \psi \bigg]^2
\Bigg\},
\label{free.energy}
\end{equation}
and the anisotropic kinetic tensor
\begin{equation}
\Lambda_{ij} = n_i n_j + \Lambda (\delta_{ij} - n_i n_j),
\label{anisotropic.diffusion.coeff}
\end{equation}
where $\epsilon = K/\xi q_0^4$ and $\Lambda = \Lambda_T / \Lambda_L$. In addition, the momentum conservation equation becomes
\begin{equation}
\partial_i P + \zeta \psi \partial_i \left( \frac{\delta F}{\delta \psi} \right) - \partial_j \sigma_{ij}^\text{D} = 0,
\label{momentum.conservation.eqn}
\end{equation}
where we have used the viscosity scale $\alpha_{4}$ to rescale the dissipative stress tensor such that
\begin{equation}
\sigma_{ij}^\text{D} = \alpha_1 n_i n_j n_k n_l v_{kl} + v_{ij} +\alpha_{56} n_k ( n_i v_{kj} + n_j v_{ki}).
\label{viscous.stress.tensor}
\end{equation}
In the rescaled momentum conservation equation a coefficient $\zeta (=\xi q_0^2/\Lambda_L \chi \alpha_4)$ appears because of the rescaling. The reader should also note that we have omitted the primes for clarity, and retained $q_0$ explicitly in the free energy, although it becomes unity in the rescaled units.

\section{Relaxation of Diblock Copolymer Lamellae}

In order to ascertain the effects of an anisotropic diffusivity and the coupling to hydrodynamics flows, we begin by investigating the linear relaxation of weakly perturbed lamellae. 
This linear analysis can be done analytically, and provides us with relaxation times that depend on the strength of the anisotropy and the hydrodynamic coupling coefficient.
Let us consider as reference state a stationary solution of the order parameter equation Eq.~(\ref{order.parameter.eqn}) for a lamellar phase, 
\begin{equation}
\psi_\text{s}({\bf r}) = \psi_1 \cos({\bf q}\cdot{\bf r}).
\label{reference.state}
\end{equation}
The wave vector ${\bf q}$ defines the unit normal to lamellae, $n_i = q_i / q$, and the kinetic constant tensor associated with the reference state is
\begin{equation}
\Lambda_{ij} = \frac{q_i q_j}{q^2} + \Lambda \left( \delta_{ij} - \frac{q_i q_j}{q^2} \right).
\end{equation}
Since $\Lambda_{ij}$ is uniform for the reference lamellae $\psi_\text{s}$, the order parameter equation, Eq.~(\ref{order.parameter.eqn}), reduces to
\begin{equation}
\partial_t \psi + v_i \partial_i \psi =
\Lambda_{ij} \partial_i \partial_j \left(\frac{\delta F}{\delta \psi}\right)
\end{equation}
or
\begin{equation}
\partial_t \psi + v_i \partial_i \psi =
\Lambda_{ij} \partial_i \partial_j \bigg[ -\epsilon \psi + \psi^3 + (\nabla^2 + q_0^2)^2 \psi \bigg].
\label{order.parameter.eqn.relaxation}
\end{equation}
We now consider small disturbances of wave number ($Q \ll q$) such that 
\begin{equation}
\begin{split}
\psi({\bf r}, t) &= \psi_1 \cos({\bf q}\cdot{\bf r}) + \psi_2(t) \exp[i({\bf q} + {\bf Q}) \cdot {\bf r}] 
\\
&
+ \psi_3(t) \exp[i({\bf q} - {\bf Q}) \cdot {\bf r}] 
+\text{c.c.} + \ldots,
\label{perturbation}
\end{split}
\end{equation}
where c.c. stands for complex conjugation, and the perturbative amplitudes $\psi_2$, $\psi_3$ and their complex conjugates are small compared to the amplitude $\psi_1$ of the reference state. We follow Zhang's study \cite{zhang10} of the effect of hydrodynamic flows on the relaxation of lamellar block copolymers to derive the amplitude equations by replacing Eq.~(\ref{perturbation}) into Eq.~(\ref{order.parameter.eqn.relaxation}) and by retaining only terms linear in $\psi_{2,3}$ and $\psi_{2,3}^{*}$.

By substituting the perturbation Eq.~(\ref{perturbation}) into Eq.~(\ref{order.parameter.eqn.relaxation}), the term in squared brackets becomes, in Fourier space (${\bf k},\omega$),
\begin{equation}
\begin{split}
\bigg[ &\frac{\delta F}{\delta \psi} \bigg] ({\bf k},\omega) 
\\
&=
\bigg[ \epsilon - (k^2 -q_0^2)^2 \bigg] \psi({\bf k},\omega) 
\\
&
-\int\frac{d^3 k_1 d\omega_1}{(2\pi)^4} \int \frac{d^3 k_2 d\omega_2}{(2\pi)^4}\; \psi({\bf k}_1,\omega_1)\psi({\bf k}_2,\omega_2) \times
\\
&
\hspace{3cm} \times \psi({\bf k}-{\bf k}_1-{\bf k}_2,\omega-\omega_1-\omega_2)
\\
&=
(2\pi)^4 M_0 \delta(\omega) \delta({\bf k}+{\bf q})
+(2\pi)^4 M_0 \delta(\omega) \delta({\bf k}-{\bf q})
\\
&
+(2\pi)^3 \bigg[ M_1(\omega) \delta({\bf k}-{\bf q}-{\bf Q})
+ M_1^{*}(\omega) \delta({\bf k}+{\bf q}+{\bf Q})
\\
& \hspace{0.8cm}
+ M_2(\omega) \delta({\bf k}-{\bf q}+{\bf Q})
+ M_2^{*}(\omega) \delta({\bf k}+{\bf q}-{\bf Q}) \bigg],
\label{elastic.part.layers}
\end{split}
\end{equation}
where
\begin{equation}
M_0 = -\frac{\psi_1}{2} \bigg[\epsilon -(q^2-q_0^2)^2 - \frac{3}{4}\psi_1^2 \bigg],
\end{equation}
\begin{eqnarray}
M_1(\omega) &=& 
-\epsilon \psi_2(\omega) + \bigg[ |{\bf q}+{\bf Q}|^2 - q_0^2 \bigg]^2 \psi_2(\omega) 
\nonumber
\\
&&
+ \frac{3}{4} \psi_1^2 \bigg[2\psi_2(\omega) + \psi_3^{*}(\omega)\bigg],
\end{eqnarray}
\begin{eqnarray}
M_2(\omega) &=& 
-\epsilon \psi_3(\omega) + \bigg[ |{\bf q}-{\bf Q}|^2 - q_0^2 \bigg]^2 \psi_3(\omega) 
\nonumber
\\
&&
+ \frac{3}{4} \psi_1^2 \bigg[2\psi_3(\omega) + \psi_2^{*}(\omega)\bigg],
\end{eqnarray}
where $\psi_{2,3}^{*}(\omega)$ are the temporal Fourier transforms of the complex conjugates of $\psi_{2,3}(t)$, respectively. 
Then we obtain the R.H.S of Eq.~(\ref{order.parameter.eqn.relaxation}) by multiplying Eq.~(\ref{elastic.part.layers}) by $-\Lambda_{ij} k_i k_j$.

Since there is no external source to generate hydrodynamic flows, the flow velocity ${\bf v}$ is produced by changes in $\psi$, and solely determined by the momentum conservation equation. In order to obtain the velocity in terms of $\psi$ we take the gradient of the momentum conservation equation Eq.~(\ref{momentum.conservation.eqn}), and solve for the pressure in Fourier space. Next, the obtained pressure is substituted back into the momentum conservation equation, resulting in
\begin{equation}
\begin{split}
v_i &({\bf k},\omega) =
\\
& - \frac{2 \zeta}{1+\alpha_{56}} \frac{1}{k^2} \left( \delta_{ij} - \frac{k_i k_j}{k^2} \right) \bigg[ \psi \partial_j \left( \frac{\delta F}{\delta \psi} \right) \bigg] ({\bf k},\omega).
\label{fluid.velocity}
\end{split}
\end{equation}
It is worth mentioning here that the flow velocity $v_i$ does not depend on the viscosity coefficient $\alpha_1$ because of the incompressibility condition. We now obtain the velocity in terms of the amplitudes of the perturbation by substituting Eqs.~(\ref{perturbation}) and (\ref{elastic.part.layers}) into Eq.~(\ref{fluid.velocity}),
\begin{equation}
\begin{split}
&v_i ({\bf k},\omega) 
\\
&=
-i(2\pi)^3\frac{2\zeta}{1+\alpha_{56}}\frac{q_j}{k^2} \left( \delta_{ij} - \frac{k_i k_j}{k^2} \right) \delta({\bf k} + {\bf Q}) \times
\\
& \hspace{0.5cm}
\times
\left[ \frac{\psi_1}{2} M_2(\omega) - \frac{\psi_1}{2} M_1^{*}(\omega) + \psi_2^{*}(\omega) M_0 - \psi_3(\omega) M_0 \right]
\\
& \hspace{0.4cm}
-i(2\pi)^3\frac{2\zeta}{1+\alpha_{56}}\frac{q_j}{k^2} \left( \delta_{ij} - \frac{k_i k_j}{k^2} \right) \delta({\bf k} - {\bf Q}) \times
\\
& \hspace{0.5cm}
\times
\left[ \frac{\psi_1}{2} M_1(\omega) - \frac{\psi_1}{2} M_2^{*}(\omega) + \psi_2(\omega) M_0 - \psi_3^{*}(\omega) M_0 \right],
\end{split}
\end{equation}
in which we have retained only $\pm {\bf Q}$ modes because the flow velocity is already linear in the perturbation amplitudes, and in the convective term of the order parameter equation it couples only to the reference state $\psi_\text{s}$ of modes $\pm {\bf q}$. Thus we find for the convective term in the order parameter equation,
\begin{equation}
\begin{split}
\int \frac{d^3 k_1 d\omega_1}{(2\pi)^4} v_i({\bf k} - {\bf k}_1,\omega-\omega_1) i k_{1  i} \psi({\bf k}_1,\omega)
\\
=
-iq_i \frac{\psi_1}{2} \bigg[ u_i ({\bf k}+{\bf q},\omega) - u_i({\bf k} - {\bf q},\omega) \bigg].
\label{convective.eqn}
\end{split}
\end{equation}

Finally combining Eqs.~(\ref{elastic.part.layers}) and (\ref{convective.eqn}) we obtain the amplitude equations, and present them separately for different modes in the small {\bf Q} limit. First there is an equation, corresponding to modes $\pm{\bf Q}$, which defines the amplitude of the reference wave $\psi_1$ 
\begin{equation}
\psi_1^2 = \frac{4}{3} \bigg[ \epsilon - (q^2 - q_0^2)^2 \bigg].
\end{equation}
It is required that $\epsilon \ge (q^2 - q_0^2)^2$ to ensure that $\psi_\text{s}$ exists.
Second, there are two amplitude equations corresponding to the modes ${\bf q} + {\bf Q}$ and $-{\bf q}+{\bf Q}$ for $\psi_2$ and $\psi_3^{*}$
\begin{equation}
\left(
\begin{array}{cc}
a & b \\
c & d
\end{array}
\right)
\left(
\begin{array}{c}
\psi_2(\omega) \\
\psi_3^{*}(\omega)
\end{array}
\right)
=0,
\label{psi2.and.psi3.eqn}
\end{equation}
where 
\begin{equation}
a = -i\omega + (H +\Lambda_{+}) l_{+} + \frac{3\psi_1^2}{4} \Lambda_{+},
\end{equation}
\begin{equation}
b = -H l_{-} + \frac{3\psi_1^2}{4} \Lambda_{+},
\end{equation}
\begin{equation}
c = -H l_{+} + \frac{3\psi_1^2}{4} \Lambda_{-},
\end{equation}
\begin{equation}
d = -i \omega + (H + \Lambda_{-}) l_{-} + \frac{3\psi_1^2}{4} \Lambda_{-}.
\end{equation}
We have introduced a hydrodynamic coupling coefficient 
\begin{equation}
H = \frac{2}{3} \frac{\zeta}{1+\alpha_{56}} \frac{1}{Q^2} \left(\delta_{ij} - \frac{Q_iQ_j}{Q^2}\right) q_i q_j \bigg[\epsilon -(q^2-q_0^2)^2\bigg],
\label{hydro.coupling}
\end{equation}
and have defined
\begin{equation}
\Lambda_\pm =
q^2 \pm 2(q_i Q_i) + \frac{(q_iQ_i)^2}{q^2} +\Lambda \Bigg[Q^2 - \frac{(q_iQ_i)^2}{q^2} \Bigg],
\label{Lambda.pm}
\end{equation}
\begin{equation}
l_\pm = (|{\bf q}\pm{\bf Q}|^2 - q_0^2)^2 - (q^2-q_0^2)^2.
\end{equation}
Additionally, there are two amplitude equations for modes $-{\bf q} - {\bf Q}$ and ${\bf q} - {\bf Q}$ leading to equatoins for for $\psi_2^{*}$ and $\psi_3$ that are the complex conjugates of Eq.~(\ref{psi2.and.psi3.eqn}). 

In general, the perturbation is linearly stable if the frequency $\omega$ is negative and pure imaginary. The imaginary frequencies are the decay rates, and they are obtained by solving the characteristic equation of the 2$\times$2 matrix in Eq.~(\ref{psi2.and.psi3.eqn}). Here we consider two different cases: an undulation mode (${\bf Q} \perp {\bf q}$) and a permeation mode (${\bf Q} \parallel {\bf q}$). When ${\bf Q} \parallel {\bf q}$, from Eqs.~(\ref{hydro.coupling}) and (\ref{Lambda.pm}) we find that neither anisotropic diffusion nor hydrodynamic coupling affect the relaxation rate. From Eq.~(\ref{psi2.and.psi3.eqn}) we find two relaxation rates
\begin{equation}
\begin{split}
\tau^{-1}_{\parallel, 1}& = \frac{3}{2} \psi_1^2 q^2
\\
& 
+ \bigg[ \frac{3}{2} \psi_1^2 + 2(11q^2 -9q_0^2)q^2 + \frac{32}{3} (q^2-q_0^2)^2 \psi_1^{-2} \bigg] Q^2 
\\
&
+ \mathcal{O}(Q^4),
\end{split}
\end{equation}
\begin{equation}
\begin{split}
\tau^{-1}_{\parallel, 2} = &
\bigg[ 2 (3q^2-q_0^2)q^2 - \frac{32}{3} (q^2 - q_0^2)^2 q^4 \psi_1^{-2} \bigg] Q^2 
\\
&
+ \mathcal{O}(Q^4),
\end{split}
\end{equation}
The relaxation rate $\tau^{-1}_{\parallel, 1}$ describes the decay of the perturbation amplitude, whereas $\tau^{-1}_{\parallel, 2}$ describes decay of its phase. Due to the order one term in $\tau^{-1}_{\parallel, 1}$ the amplitude decays much faster than the phase, and follows adiabatically any change in the phase. 
%
%

When ${\bf Q} \perp {\bf q}$, we have $l_{\pm} = 2(q^2-q_0^2) Q^2 + Q^4$, and $\Lambda_{\pm}=q^2 + \Lambda Q^2$. In this case Eq.~(\ref{psi2.and.psi3.eqn}) reduces to
\begin{equation}
\left(
\begin{array}{cc}
a_\perp & b_\perp \\
b_\perp & a_\perp
\end{array}
\right)
\left(
\begin{array}{c}
\psi_2(\omega) \\
\psi_3^{*}(\omega)
\end{array}
\right)
=0,
\label{psi2.and.psi3.eqn.perpendicular}
\end{equation}
where
\begin{equation}
a_\perp = -i\omega + (H +\Lambda_{+}) l_{+} + \frac{3\psi_1^2}{4} \Lambda_{+},
\end{equation}
\begin{equation}
b_\perp = -H l_{+} + \frac{3\psi_1^2}{4} \Lambda_{+}.
\end{equation}
Then it is straightforward to obtain two relaxation rates
\begin{equation}
\tau^{-1}_{\perp, 1} =
\frac{3}{2} \psi_1^2 q^2 + \bigg[ 2(q^2 -q_0^2)q^2 + \frac{3}{2} \psi_1^2 \Lambda \bigg] Q^2 + \mathcal{O}(Q^4),
\end{equation}
\begin{equation}
\begin{split}
\tau^{-1}_{\perp, 2} =& 4h(q^2-q_0^2) + 2 \bigg[ h + (q^2-q_0^2)q^2 \bigg] Q^2
\\
&+\bigg[ q^2 + 2\Lambda(q^2 - q_0^2) \bigg]Q^4 + \mathcal{O}(Q^6),
\end{split}
\end{equation}
where 
\begin{equation}
h = \frac{2}{3} \frac{\zeta}{1+\alpha_{56}} q^2 \bigg[\epsilon -(q^2-q_0^2)^2\bigg].
\label{hydro.coupling.1}
\end{equation}
Again these two rates $\tau^{-1}_{\perp, 1}$ and $\tau^{-1}_{\perp, 2}$ govern the relaxation of the amplitude and phase, respectively.
We find that anisotropic diffusion contributes at order $Q^2$ to $\tau_{\perp,1}^{-1}$, and at order $Q^4$ to $\tau_{\perp,2}^{-1}$; therefore, its effect on the relaxation of weakly perturbed lamellae is negligible in the limit of small $Q$. Hydrodynamic flow only couples to the phase of perturbation,
%
%
and 
derives the phase to decay faster for $q\neq q_0$ unlike the previous case of parallel perturbations. If $q = q_0$, from Eq.~(\ref{hydro.coupling.1}) we can define the hydrodynamic diffusion length $\lambda_\text{hydro}$ as
\begin{equation}
\lambda_\text{hydro}^2 \equiv \frac{1}{hq_0^2} = \frac{3}{2} \frac{\Lambda_\text{L} (\alpha_4 + \alpha_{56})\chi}{K},
\end{equation}
in the dimesional units.
Then, the marginal mode for instability, the diffusion due to the hydrodynamic flow becomes negligible with respect to the isotropic order parameter diffusion when $\lambda_\text{hydro} Q > \sqrt{2}$.
In this case the diffusion mode becomes identical to the undulation mode of Smectic-A liquid crystal ($\sim Q^4$).\cite{deGennes.liquid.crystals.93} When $\lambda_\text{hydro} Q < \sqrt{2}$, the hydrodynamic diffusion dominates with a decay rate proportional to $Q^2$.
%
%

\section{Grain Boundary Motion}

\begin{figure}[t]
\centering
   \includegraphics[width=7cm]{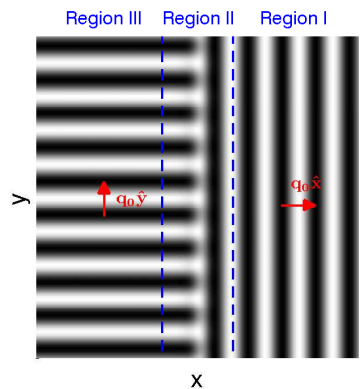}
  \caption{Sketch of a 90$^\circ$ tilt grain boundary separating two semi-infinte lamellae. In regions I and III the lammellae are nearly in their stationary state with wavevectors $q_0 \hat{x}$ and $q_0 \hat{y}$, respectively. The grain boundary region (II) is the region between the two blue dashed lines. In it, both envelopes $A$ and $B$ exhibit large variation.}
  \label{GB}
\end{figure}

We have studied in the previous section the effect that an anisotropic diffusion tensor can have on the linear relaxation dynamics of a weakly perturned lamellar phase. In general, however, when a block copolymer is brought below its microphase separation transition point, a large number of structural defects such as grain boundaries are quenched in a spatially extended system. These boundaries separate locally layered domains of different orientations producing a macroscopically inhomogeneous system. Within the defected region (of extent which is much larger that the lamellar wavelength in the weak segregation limit considered in this paper) undulation and permeation diffusive modes strongly couple. This coupling leads to a siginificant contribution to defect dynamics, the subject matter of this section.

Grain boundary motion in layered systems has been intensively studied in other contexts such as Rayleigh-B\'enard convection \cite{manneville83, tesauro87, boyer01}. In the governing equation for Rayleigh-B\'enard convective rolls, the order parameter is the vertical velocity which is equivalent to the order parameter in equation Eq.~(\ref{order.parameter.eqn}). In this section we examine the consequences of allowing an anisotropic kinetic coefficient, compatible with the uniaxial symmetry of a layered block copolymers. Our results extend the analysis of Rayleigh-B\'enard convective rolls given in Refs.~\cite{manneville83, tesauro87, boyer01}.

We consider a 90$^\circ$ tilt grain boundary that separates two semi-infinite domains of block copolymer lamellae with wavevector $q_0 \hat{\bf x}$ in region I, and $q_0 \hat{\bf y}$ in region III as shown in Fig.~\ref{GB}. For simplicity, we focus on an effective two-dimensional system by taking advantage of translational symmetry of lamellae along the direction ($\hat{\bf z}$), perpendicular to the wavevectors of the two semi-infinite lamellae.

In order to take into account the inhomogeneous nature of the kinetic constant tensor $\Lambda_{ij}$ due to the presence of a 90$^{\circ}$ grain boundary in the interface region II, we model $\Lambda_{ij}$ with an auxiliary function $\Theta(x)$ that interpolates smoothly within the width of the grain boundary from unity in the region where one semi-infinite lamella is present to zero in the opposite region:
\begin{eqnarray}
\Lambda_{ij} &=&
\Theta(x) \Lambda_{ij}^\text{I} + \Theta(-x) \Lambda_{ij}^\text{III}
\nonumber
\\
&=&
\Theta(x) \bigg[ \Lambda \delta_{ij} + (1-\Lambda) \delta_{ix} \delta_{jx} \bigg]
\nonumber
\\
&&
+ \Theta(-x) \bigg[ \Lambda \delta_{ij} + (1-\Lambda) \delta_{iy} \delta_{jy} \bigg],
\label{GB.Lambda}
\end{eqnarray}
where $\Lambda_{ij}^\text{I}$ and  $\Lambda_{ij}^\text{III}$ are the kinetic constant tensors defined in regions I and III with $q_0\hat{x}$ and $q_0\hat{y}$, respectively, and $\Theta(-x) = 1-\Theta(x)$.
In writing $\Lambda_{ij}$ as Eq.~(\ref{GB.Lambda}) we have assumed that the two semi-infinite lamellae decay with a common length scale proportional to the size of grain boundary. In Refs.~\cite{manneville83, tesauro87} it is found that the width of grain boundary diverges as $\epsilon^{-1/2}$ in the weak segregation limit, so that we can infer that $\Lambda_{ij}$ changes very slowly in the interfacial region.

Due to the inhomogeneity of $\Lambda_{ij}$ its gradient does not vanish, and the order parameter equation becomes
\begin{equation}
\begin{split}
\partial_t \psi =
\bigg\{ (1-\Lambda) [\partial_x \Theta(x)] \partial_x + \Lambda_{ij} \partial_i \partial_j \bigg\} \times
\\
\times\bigg[ -\epsilon \psi + \psi^3 + (\nabla^2 + q_0^2)^2 \psi \bigg],
\end{split}
\end{equation}
where the coupling to hydrodynamic flow is neglected for the analysis in this section. We now use a multiple scale analysis to derive amplitude equations close to the linear instability threshold ($\epsilon \ll 1$). Following Tesauro and Cross \cite{tesauro87}, we introduce slow variables $\bar{X} = \epsilon^{1/4} x$, $X = \epsilon^{1/2} x$, $Y = \epsilon^{1/4} y$, $\bar{Y} = \epsilon^{1/2} y$, $T = \epsilon t$, and expand the derivatives $\partial_x \to \partial_x +\epsilon^{1/4} \partial_{\bar{X}} + \epsilon^{1/2} \partial_{X}$, $\partial_y \to \partial_y + \epsilon^{1/4} \partial_{Y} + \epsilon^{1/2} \partial_{\bar{Y}}$, and $\partial_t \to \epsilon \partial_{T}$. Next, by noting that the order parameter scales as $\epsilon^{1/2}$ we take
\begin{equation}
\psi = \frac{\epsilon^{1/2}}{\sqrt{3}} \bigg[ A \exp (i q_0 x) + B \exp (i q_0 y) \bigg]+\text{c.c.},
\end{equation}
where $A$ and $B$ are functions of the slow variables only.
Then the amplitude equations at $\mathcal{O}(\epsilon^{3/2})$ are,
\begin{equation}
\partial_T A = -q_0^2 \frac{\delta F^\prime}{\delta A^{*}} + \Delta \bigg[ i q_0 \partial_x \Theta(x) + q_0^2 \Theta(-x) \bigg]
\frac{\delta F^\prime}{\delta A^{*}},
\label{eqn.A}
\end{equation}
\begin{equation}
\partial_T B = - q_0^2 \frac{\delta F^\prime}{\delta B^{*}} + \Delta q_0^2 \Theta(x) \frac{\delta F^\prime}{\delta B^{*}},
\label{eqn.B}
\end{equation}
where the free energy functional is
\begin{equation}
\begin{split}
&F^\prime [A,A^{*},B,B^{*}] 
\\
&= \int d^2 r \Bigg\{
- |A|^2 - |B|^2 + \frac{1}{2} \bigg( |A|^4 + |B|^4 \bigg) + 2 |A|^2|B|^2
\\
& \hspace{1cm}
+ \bigg| (2iq_0 \partial_x  + \partial_y^2) A \bigg|^2
+ \bigg| (2iq_0 \partial_y  + \partial_x^2) B \bigg|^2
\Bigg\},
\end{split}
\end{equation}
and $\Delta = 1 - \Lambda$. Since $\Lambda$ is positive, 
it is required that $\Delta<1$. It is easy to show that the isotropic case of dissipative coefficient ($\Lambda_{ij} \sim \delta_{ij})$ is recovered by taking $\Lambda =1$. 

The solutions $A^\text{s}$ and $B^\text{s}$ of the stationary planar 90$^\circ$ grain boundary without anisotropy, given in Refs.~\cite{manneville83} and \cite{tesauro87}, are the stationary solutions of the amplitude equations. Both $A^\text{s}$ and $B^\text{s}$ saturate to $\epsilon^{1/2}$ as $x$ tends to $+\infty$ or $-\infty$ respectively, but have different decaying behaviors within the grain boundary. The amplitude $A^\text{s}$ has a longer decaying length scale $\sim \epsilon^{-1/2}$ than $B^\text{s}$. Since the width of the grain boundary scales as $\epsilon^{-1/2}$, it is reasonable to assume that this is the same scale of variation of the function $\Theta(x)$. This leads to $\partial_i \Lambda_{ij} \sim \epsilon^{1/2}$, and its contribution to the amplitude equations appears at higher order in $\epsilon$. The remaining relevant term is $\Lambda_{ij} \partial_i \partial_j (\delta F/\delta \psi)$, and Eqs.~(\ref{eqn.A}) and (\ref{eqn.B}) reduce to
\begin{equation}
\partial_T A = - q_0^2 \bigg[ 1 - \Delta \Theta(-x) \bigg]
\frac{\delta F^\prime}{\delta A^{*}},
\label{eqn.A.1}
\end{equation}
\begin{equation}
\partial_T B = - q_0^2 \bigg[ 1 - \Delta \Theta(x) \bigg] \frac{\delta F^\prime}{\delta B^{*}}.
\label{eqn.B.1}
\end{equation}

We now use the energy method \cite{manneville83, cross93} to calculate the grain boundary velocity. The time derivative of $F^\prime$ is given by
\begin{eqnarray}
\frac{d F^{\prime}}{d t} &=& 
- \frac{2}{q_0^2} \int d^2 r  
\frac{1}{1-\Delta \Theta(-x)} |\partial_t A|^2
\nonumber
\\
&&
- \frac{2}{q_0^2} \int d^2 r  
\frac{1}{1-\Delta \Theta(x)} |\partial_t B|^2.
\end{eqnarray}
Note that the anisotropy effect is on the R.H.S. only, and that the coefficients of both $|\partial_t A|^2$ and $|\partial_t B|^2$ are positive because $\Delta < 1$. When the grain boundary moves with a velocity $v_\text{GB}$, it is convenient to take $A^\text{s}(x-v_\text{BG}t)$ and $B^\text{s}(x-v_\text{BG}t)$ so that they are stationary in the moving frame. Then the time derivative can be replaced $\partial_t$ with $-v_{BG} \partial_x$, and we find for the grain boundary velocity
\begin{equation}
v_\text{GB} = M \int dy \bigg[ \mathcal{F}^\prime( x=\infty, y) - \mathcal{F}^\prime(x=-\infty, y) \bigg],
\label{GB.velocity}
\end{equation}
where $\mathcal{F}^\prime$ is the free energy density, and the effective mobility of the boundary is given by
\begin{eqnarray}
\frac{1}{M} &=&
\frac{2}{q_0^2} \int d^2 r  
\frac{1}{1-\Delta \Theta(-x)} |\partial_x A^\text{s}|^2
\nonumber
\\
&&
+\frac{2}{q_0^2} \int d^2 r  
\frac{1}{1-\Delta \Theta(x)} |\partial_x B^\text{s}|^2.
\label{eq:mobility}
\end{eqnarray}
Since $\Delta$ is an order one quantity, the contribution to the boundary velocity due to anisotropic diffusion is large. The envelope $A$ relaxes in region I (of dominant orientation $q_0\hat{x}$) differently than in region III (of dominant orientation $q_0\hat{y}$). In region I, diffusion is along the lamellar normal, whereas this component of the order parameter evolves through transverse difussion in region III. Exactly the same is true of component $B$. As a consequence, the boundary velocity depends on a weighted average of the two independent diffusion coefficients, with the weight function being the gradient of the order parameter envelopes, as given in Eq. (\ref{eq:mobility}). Of course, a similar qualitative behavior can be expected in the vicinity of other structural defects.
When $\Delta=0$, the isotropic result of Refs.~\cite{manneville83, boyer01} is recovered. 

We note that there is no grain boundary motion for unperturbed lamellae when $\mathcal{F}^\prime(x=\infty) = \mathcal{F}^\prime(x=-\infty) \sim \epsilon$. In practice, an imbalance of the free energies caused by external sources is necessary to drive grain boundary motion so as to reduce excess free energy. We find that anisotropy enhances (reduces) $v_\text{GB}$ when $\Delta<0$ ($0<\Delta<1$).

\section{Conclusion}

We have investigated diffusive relaxation in lamellar phases of block copolymers when allowing for uniaxial symmetry of the consitutive law between diffusive forces and fluxes, as well as hydrodynamic coupling. We have shown that coupling to flows leads to a relaxation rate proportional to $Q^{2}$, where $Q$ is the wavenumber of the characteristic perturbation. The uniaxial symmetry of the lamellar phase of a diblock copolymers requires an anisotropic kinetic constant in the order parameter equation, and an anisotropic stress tensor in the momentum conservation equation. With them, we have calculated the relaxation rates of a weakly perturbed lamella, and found that the effect of anisotropy becomes negligible compared to either hydrodynamic flow or (isotropic) order parameter diffusion.
We have also studied the motion of a grain boundary by calculating its velocity, and shown that the velocity is significantly affected by anisotropic diffusion.


\section*{Acknowledgments}
We thank F. Drolet for valuable discussions and the Minnesota Supercomputing Institute for support.

\footnotesize{
\bibliography{block.copolymers} 
\bibliographystyle{rsc} 
}

\end{document}